\documentclass[a4paper,11pt]{article}
\pdfoutput=1 

\usepackage{jcappub} 

\usepackage[T1]{fontenc} 
\usepackage{subfigure}
\usepackage{multirow}

\newcommand{\mpl}{M_{\rm Pl}}
\newcommand{\vtilde}{V}
\newcommand{\ttilde}{T}
\usepackage{amsmath}

\DeclareMathOperator\sech{sech}
\DeclareMathOperator\csch{csch}
\def \beq{\begin{equation}}
\def \eeq{\end{equation}}

\title{\boldmath Theoretical and observational constraints on Tachyon Inflation}

\author[a]{Nandinii Barbosa-Cendejas,}
\author[b,c]{Josue De-Santiago,}
\author[b]{Gabriel German,}
\author[b]{Juan Carlos Hidalgo,}
\author[d]{Refugio Rigel Mora-Luna.}

\affiliation[a]{Facultad de Ingenier\'ia El\'ectrica,
Universidad Michoacana de San Nicol\'as de Hidalgo,\\Morelia, Michoac\'an, M\'exico}
\affiliation[b]{Instituto de Ciencias F\'{\i}sicas, Universidad Nacional Aut\'onoma de M\'exico,\\Apdo. Postal 48-3, 62251 Cuernavaca, Morelos, M\'exico}
\affiliation[c]{Instituto de Ciencias Nucleares, Universidad Nacional Aut\'onoma de M\'exico,\\Apdo. Postal 70-543, Ciudad de M\'exico, M\'exico}
\affiliation[d]{Facultad de Ingenier\'ia Qu\'{i}mica,
Universidad Michoacana de San Nicol\'as de Hidalgo,\\Morelia, Michoac\'an, M\'exico}

\emailAdd{nandini@fis.unam.mx}
\emailAdd{josue@ciencias.unam.mx}
\emailAdd{gabriel@fis.unam.mx}
\emailAdd{hidalgo@fis.unam.mx}
\emailAdd{rigel@fis.unam.mx}

\abstract{We constrain several models in Tachyonic
  Inflation derived from the large-$N$ formalism by considering
  theoretical aspects as well as the latest observational
  data. On the theoretical side, we
  assess the field range of our models by means of the excursion of the
  equivalent canonical field. On the observational side, we employ
  BK14+PLANCK+BAO data to perform a parameter
  estimation analysis as well as a Bayesian model selection to distinguish
  the most favoured models among all four classes here presented. We
  observe that the original potential $V \propto \sech(T)$  is strongly disfavoured by observations with respect to a reference model with 
  flat priors on inflationary observables. This realisation of Tachyon inflation also presents a large field range which may demand further quantum corrections. We also provide examples of
  potentials derived from the polynomial and the perturbative classes which are
  both
  statistically favoured and theoretically acceptable. 
}
\begin{document}

\maketitle
\flushbottom

\section{Introduction}
\label{sec:intro}

Inflation is a solid proposal to solve some of the Hot Big Bang cosmological 
problems, however the nature of the inflationary mechanism is not yet 
elucidated and a large number of models have been proposed. 
The diversity of models for inflation are classified in different scenarios
where, for the case of a single field, the inflation is driven by a
canonical scalar field \cite{inflation}, or a field of non-canonical nature
\cite{non-canonical}. Some
of the more attractive models have been motivated by high energy theories and in particular
tachyon inflation has been proposed as a realization of type-IIA and type-IIB
string theories \cite{sen:todas}. In order to explore the possibilities of the tachyon
inflationary model, recent studies  have presented diverse
realizations and constraints to the tachyon potential
\cite{Daniel,tachyon1,Li:2013cem,otrostaquiones}. In
particular \cite{Nan} presents  several universality classes, derived
using the large-$N$ formalism \cite{nformalism,Fei:2017fub,Garcia-Bellido:2014wfa}. While many of these new realizations
of the tachyon potential reproduce the observed values of the spectral index
and satisfy the observational bounds to the tensor perturbations for a given number
of e-folds $N$, 
it is desirable to perform a full Bayesian analysis of the models in
light of the recent observations to determine
which models are favoured by the data. On the theoretical side it is
important to have under control the quantum corrections to the excursion of
the field during inflation, in particular if $N$ is large \cite{Baumann:2009ds}.

In this paper we combine the latest observational data from the CMB anisotropies
\cite{Adam:2015rua,Array:2015xqh} and
the BAO signal \cite{baotodos} to perform a parameter estimation of
the families of models within each class, as well as a
Bayesian evidence analysis to distinguish which universality classes
and the specific models favoured by the data. We look at the
canonical field models associated to each of the tachyon potentials in
order to estimate the field excursion $\Delta \phi$.
The models with a trans-planckian excursion in $\Delta \phi$ require a complete treatment of
their original string model in order to ensure that they are dynamically controlled
over a trans-planckian field range (see e. g. \cite{Baumann:2009ds}).

The paper is organized as follows: In section \ref{Sec:II} we present the relevant equations
of tachyon inflation and its associated field excursion in terms of the tachyon and
the canonical fields.
In section \ref{Sec:III} we present, in the context of the large-$N$ formalism,
the universality classes of tachyon inflation and the expressions for their
observational parameters. For each one of these classes we perform a parameter estimation
analysis in order to obtain the preferred models within each class. In this
section we also present the canonical field
excursion for each of the classes. In section \ref{Sec:modelselection} we
compute the Bayes factor for each universality class in order to assess their relative
probability and propose a set of preferred models for tachyon inflation. In
section \ref{Sec:V} we analyse our results, together with the theoretical
criteria of a sub-planckian excursion of the associated canonical
field. We finally put forward the most favoured models of tachyon inflation.

\section{Field excursion in the large-$N$ formalism}
\label{Sec:II}

In order to quantify the expansion of space during the inflationary
period, it is useful to define the number of e-folds to the end of inflation given by $\Delta N=\log(a_e/a_*)$ where $a_e$ corresponds to the scale factor at the end of inflation and $a_*$ is the scale factor at the time when the largest observable scales exit the horizon. This means that
$N$ will change from $N_*\sim 10^2$ at the beginning of (the observable part of)
inflation to $N_e\sim 0$ towards the end of the inflationary period. 
The typical figure for $\Delta N$ necessary to solve the Big Bang problems is around $60$. Since a variety
of scenarios can account for the sustained accelerated expansion, the
large-$N$ formalism was developed as an attempt to parametrize the
observables of inflation in models with the common characteristic of large $N$ values.
In this parametrization, the (small) slow-roll parameters are prescribed as functions of (large) $N$. Consequently, important observables of the model $(n_s,r)$, and their running, are all functions of $N$. 
An analysis of different inflationary models by this method and the contrast against recent data is presented for example, in
\cite{Nan,nformalism,Fei:2017fub,Garcia-Bellido:2014wfa}.
In particular, an important quantity that can be computed within this formalism is the interval that the field covers during inflation. In reference \cite{Garcia-Bellido:2014wfa}, the excursion of the field $\Delta \phi$ is computed for the classes covered by the large-$N$ formalism applied to canonical field inflation. This excursion must not exceed the Planck scale $M_{\rm Pl}$ in order to avoid quantum corrections to the classical treatment of inflation. Here we test tachyon inflation with the same criterion, by looking at the classes of potentials derived within the Large-$N$ formalism \cite{Nan}.

The Tachyon Scalar Field Inflation (TSFI) scenario was proposed by Sen in \cite{sen:todas} and also studied in \cite{otrostaquiones}
as well as in thick brane world models like the ones in \cite{alfredo}
with an effective action of the tachyonic condensate $T$ given by the expression
\begin{equation}\label{eq:action}
   S=\int d^4x \sqrt{-g} V(T) \sqrt{1 - \partial_\mu T \partial^\mu T}\,,
\end{equation}
where $T$ has dimensions of $\mpl^{-1}$.
With this prescription the energy density of the field is obtained as
\begin{equation}
   \rho = \frac{V(T)}{\sqrt{1-\partial_\mu T \partial^\mu T}}\,.
\end{equation}
For a flat Friedmann-Robertson-Walker metric,
the Raychaudhuri equation can be used to show the way the tachyon field drives the accelerated 
expansion of the universe, in this case it reads
\begin{eqnarray}\label{ace:condition}
\frac{\ddot a}{a} &=& -\frac{1}{6 M^{2}_{Pl}}(\rho + 3P)
\nonumber \\ &=&
\frac{1}{3 M^{2}_{Pl}}\frac{\vtilde}{\sqrt{1-\dot \ttilde^{2}}}\left(1-\frac{3}{2}\dot \ttilde^2
\right) \,.
\end{eqnarray}

The accelerated expansion is controlled by the slow-roll parameters. We employ the "Hubble flow" slow-roll parameters \cite{Schwarz:2001vv}\footnote{also motivated by the renormalization group description of the inflationary evolution \cite{Binetruy:2014zya}}, defined by 
\begin{eqnarray}
\epsilon_{i+1} \equiv d \log |\epsilon_i| / dN\,, \quad \mathrm{for}\,\, i \ge 0\quad 
\mathrm{and} \quad \epsilon_0 \equiv H.
 \end{eqnarray}
 
\noindent For the TSFI scenario, the first Hubble parameter presents a simple relation given by
\begin{equation}
    \epsilon_{1}=\frac{3}{2}\dot \ttilde^{2}.
\end{equation}

\noindent From Eq.~\eqref{ace:condition} it is clear that accelerated expansion is guaranteed for $\epsilon_1 < 1 $. 

In order to determine the excursion of the tachyon field $\Delta \ttilde$ in terms of the number of
$e$-folds we consider the definition of the parameter $\epsilon_1$ and write
\begin{equation}\label{eq:dT}
   \Delta N = \sqrt{\frac{3}{2}}\int^{T_{\rm e}}_{T_*} \frac{H}{\sqrt{\epsilon_1}}dT\,.
\end{equation}
Recalling that the slow-roll parameter is related to the tensor-to-scalar ratio $r=16\epsilon_1$, the above integral can be expressed in terms of observables alone. This 
together with the COBE normalization \cite{tachyon1}
\begin{equation}\label{COBE}
H_{*}= r^{1/2} 1.0312  \times 10^{-4} M_{\rm Pl} \,,
\end{equation}

\noindent can be used to obtain an estimation for the field excursion as
\begin{equation}\label{firstbound}
\Delta \ttilde \simeq \frac{1980}{M_{\rm Pl}}\Delta N \,.
\end{equation}

\noindent where we take the approximation that $\epsilon_1/H^2$ is constant (this has been derived also
in ref. \cite{Fei:2017fub}).
It is interesting to note that, unlike the Lyth bound \cite{Lyth:1996im}
for a canonical scalar field, where the excursion is
a function of the amplitude of tensor perturbations $r$ (and therefore
is still an unknown quantity), the tachyon
excursion depends exclusively on the COBE normalization and the number of e-folds
of inflation. Therefore, in the tachyon case, the excursion is always known
at least to a first approximation.

Due to the units of the tachyon field, however, the eventual quantum corrections to the classical analysis are not directly assessed by its excursion. In the literature, the limit of the classical analysis  is assessed through the excursion of the canonical field. Indeed, quantum corrections become important if the excursion of the canonical field is larger than the Planck mass \cite{Lyth:1996im,Lidsey:1995np}. A tachyon field scenario has been mapped to its corresponding canonical field in \cite{macorra}.
The field excursion for the canonical field corresponding to the TSFI scenario is given by
integrating $\dot \phi = \sqrt{V} \dot T$. That is,
\begin{equation}
   \Delta \phi = \int_{T_*}^{T_{e}} \sqrt{V(T)} dT \,.
   \label{eq:canonical_transformation}
\end{equation}

\noindent This is valid when the time derivative of the tachyon field $\ttilde$ is small compared
to one, which happens during inflation.

In the following section  we will obtain more accurate numbers for $\Delta \ttilde$ (and consequently $\Delta \phi$)
using the large-$N$ formalism for the most representative universality classes
that comprise different behaviours of $V(T)$, keeping in mind that the limit $\Delta \phi>\mpl$ is an undesirable feature of the inflationary theory.

\section{Constraints for universality classes}
\label{Sec:III}
In this section we will present an in depth analysis for each of the universality
classes obtained in Ref. \cite{Nan} which correspond to different parametrizations of the
Hubble flow parameters for the TSFI, that in turn translates to
models with different potentials $V(T)$. For each of these classes we will perform a
Bayesian parameter estimation using CMB and BAO observations. Then
we will evaluate the field excursion of both the tachyon and the
associated canonical field during the period of inflation.
The parameter estimation of this section only tells us which is the most
preferred model (potential) within each class given the observations, but does not provide
a comparison between classes. In section \ref{Sec:modelselection} we
carry out an analysis of the Bayesian evidence in order to 
find out which class is preferred  according to the observations.

In order to obtain these results we
used two different sets of data; for the first dataset, we take CMB polarization data from the 2015 data release of the
Planck satellite \cite{Adam:2015rua}, while the second dataset takes CMB polarization data from BICEP-Keck 2014 (BK14) \cite{Array:2015xqh}. In both cases we have used temperature data from Planck 2015 and BAO measurements from SDSS \cite{baotodos}. In order to analyze the posterior
distribution functions, we ran a Metropolis algorithm with 7 cosmological parameters where
three of them are related to the inflationary scenario ($\log(A_s)$, $N_*$,$\lambda$). The other four parameters are related to the later evolution of the universe($\theta_{MC}$, $\Omega_b h^2$, $\Omega_c h^2$,$\tau$) but their posterior probabilities are correlated to those of the inflationary parameters. Thus, while we are mainly interested on the posterior distributions of the former set, the latter are needed to adequately analyse the CMB data.

We imposed flat priors to all the parameters; in particular we chose $\lambda$ to be positive, which is required to have $\epsilon_1$ positive. As for the priors for $\Delta N$, from Ref.  \cite{Lyth:2009zz}
the lower limit $N_{*\mathrm{min}} = 15$ is required to have inflation at an energy scale at least higher than nucleosyntesis.
On the other hand, for $\Delta N$ larger than 61 the evolution of the universe requires a period
of kination where a stiff fluid dominates the energy density before the onset of nucleosynthesis. Here we loosen that limit and use $ \Delta N_{\mathrm{max}}=75$ (a smaller number than that of \cite{Liddle:2003as}, where $\Delta N_{\mathrm{max}}=83.5$).
For practical purposes we used these limits on $\Delta N$ as valid for $N_*$ under the assumption that $N_e \ll 1$.
This is valid for the perturbative, hyperbolic secant and polynomial classes (derived below), where 
$\epsilon_1(N_e)\sim 1$ for $N_e \sim 0$. The other studied case, the exponential class, presents potentials that do not reach the $\epsilon\to 1$ limit. This will be discussed with more detail below.

In order to analyze each of the universality classes we use the expressions obtained in
our previous work \cite{Nan} which we summarize in the following lines.
The formalism consists of establishing an a-priori relation between the Hubble flow parameter $\epsilon_1$ and the number of e-folds
$N$ left to the end of inflation. With this information one can compute the potential and the tachyon field
as functions of $N$ by means of the relations:
\begin{eqnarray}
   \vtilde(N) &=&  V_0 \exp\left[ \int 2\epsilon_1(N) dN \right] \,,\nonumber \\
   \ttilde(N) &=&  \mpl \int \sqrt{\frac{2 \epsilon_1(N)}{ \vtilde (N)}} dN \,.
   \label{classes_basic}
\end{eqnarray}
From the above relations it is also possible to solve for $V(T)$. Finally,
it is possible to eliminate the dependence on $V_0$ of the tachyon field by using
the COBE normalization (\ref{COBE}) that can be expressed as
\begin{equation}
   \vtilde(N_*) = 3.38\times 10^{-8}\mpl^4 r \,.
   \label{}
\end{equation}

In order to compare the predictions of the different classes to the observations we employ the well-known relations between the Hubble flow parameters and the tensor-to-scalar ratio $r$ and the spectral index $n_s$ \cite{Schwarz:2001vv}
\begin{equation}
r = 16\epsilon_1, \qquad n_s = 1 - 2 \epsilon_1 - \epsilon_2.
\end{equation}
 
\noindent In the following subsections we will apply all of the above considerations to each one of
the classes obtained in Ref. \cite{Nan}.

\begin{figure}[h]
\begin{center}
\subfigure[Inflation parameters for the perturbative class with potential given by (\ref{eq:perV}). \hskip0.1cm The potential $T^2$
   corresponds to $\lambda=1/4$ which is discarded at $2\sigma$ by the BK14 observations.]{\label{fig:pert}
\includegraphics[width=6cm]{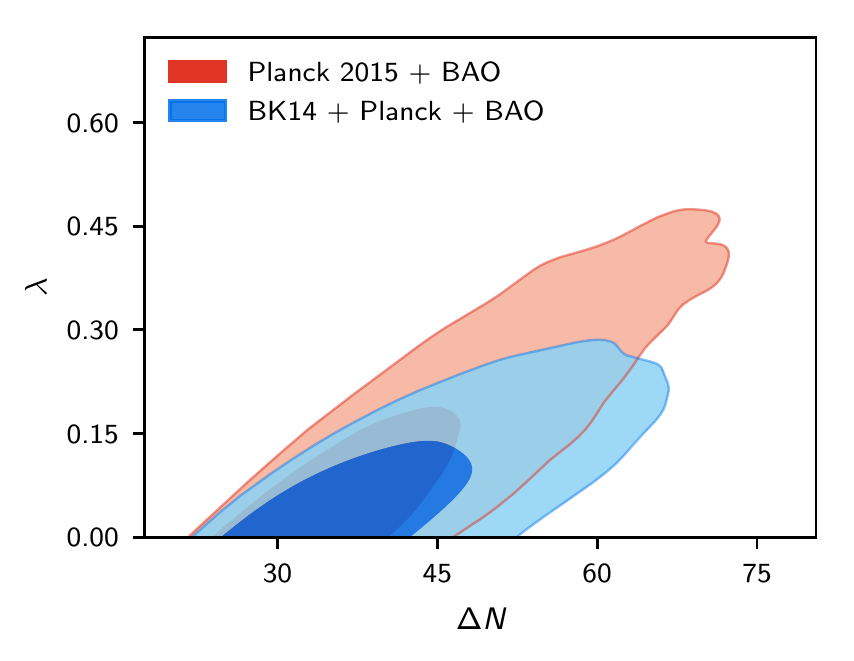}}
\subfigure[Inflation parameters for the polinomial class with potential given by eq. (\ref{pot_pol}).]{\label{fig:pol}
\includegraphics[width=6cm]{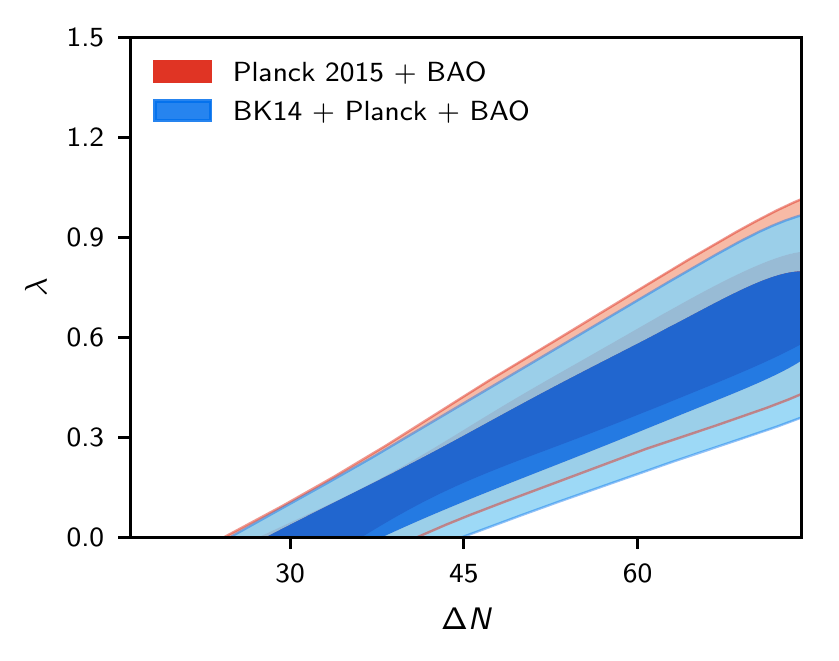}}
\subfigure[Inflation parameters for the exponetial class with potential given by (\ref{pot_expfirst}).]{\label{fig:exp}
\includegraphics[width=6cm]{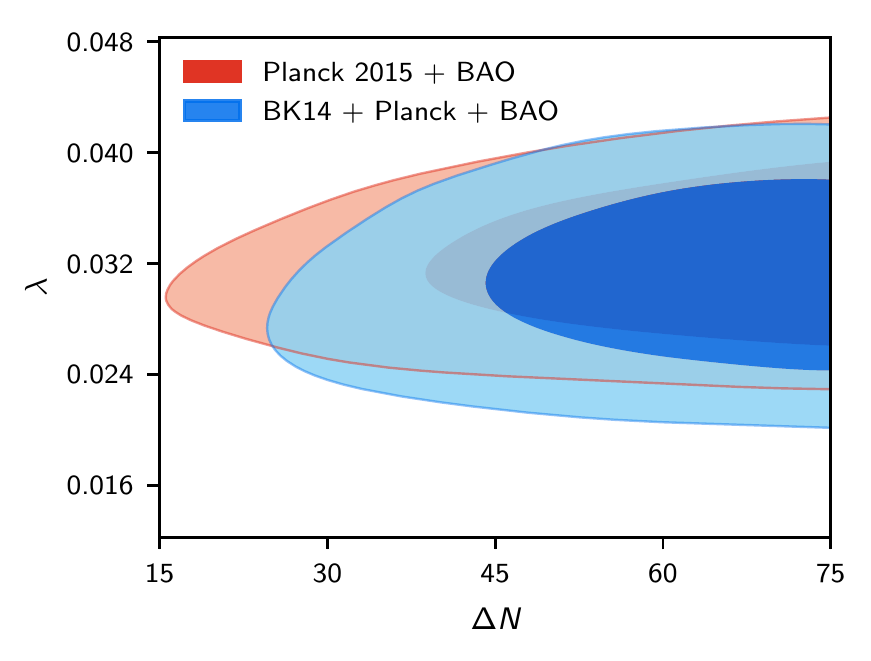}}
\subfigure[Inflation parameters for the sech potential given by eq. (\ref{pot_sech}).]{\label{fig:sech}
\includegraphics[width=6cm]{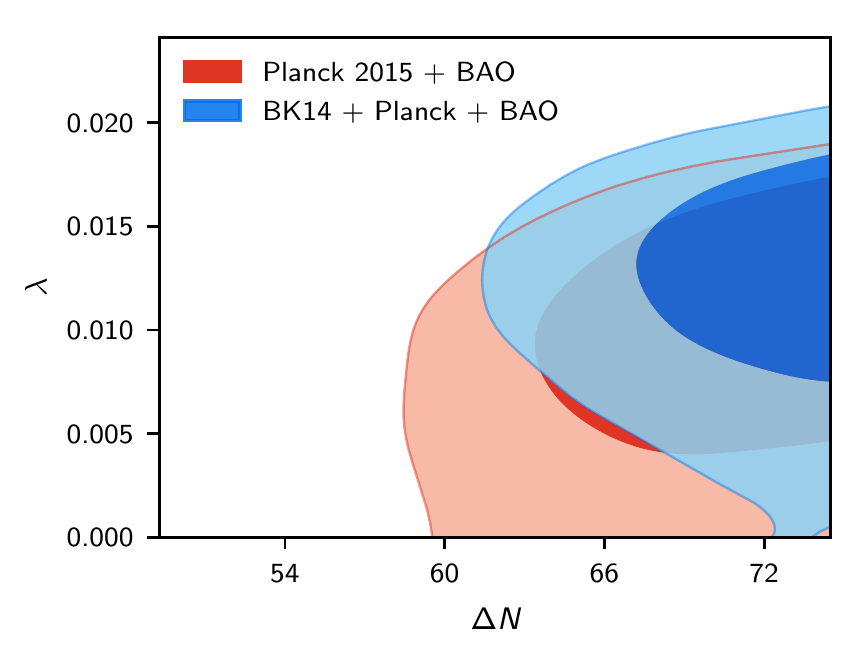}}
\end{center}
\caption{Marginalized confidence regions at 68\% and 95\% (within the displayed fraction of parameter space) for the analyzed universality classes. The posterior distributions were obtained with flat priors for $\lambda$ and $N^{*}$ and with limits discussed in the text.
The parameter space was sampled using the Metropolis Hastings algorithm implemented in CosmoMC \cite{Lewis:2002ah}, and CAMB \cite{Lewis:1999bs} as the Boltzmann solver. 
Note that the posterior distribution of the parameters $\left\{\lambda,\Delta N \right\}$ is restricted to an acceptable range of $\Delta N$ (see text). 
This may not include the most likely values of the parameters given the data. In each case this range avoids super-planckian densities and exotic matter components dominating the early universe after inflation.}
\label{fig:obs_nlam}
\end{figure}
\subsection{Perturbative class}
\label{Sec:IIIa}
This class is parametrized by the Hubble flow parameter 
\begin{equation}
\epsilon_1 =\frac{ \lambda}{ N} \,,
\end{equation}
where $\lambda$ is a positive constant that defines the particular model within the class.
The inflation period ends when this parameter reaches the value of one, corresponding to $N_e=\lambda$.
As we showed in Ref. \cite{Nan} the potential for this class becomes
$V=V_{0}N^{2 \lambda}$ which, in terms of the tachyon field $\ttilde$
corresponds to 
 \begin{equation} \label{eq:perV}
  \vtilde =  \begin{cases}
    \hat V_{0}  \ttilde^n & 0<\lambda<\frac{1}{2} \,, \\
    V_0 \,\text{exp}\left( \frac{\sqrt{V_0} \ttilde}{\mpl} \right)
    & \lambda= \frac{1}{2} \,,\\
  \end{cases}
\end{equation}
where $n=4\lambda/(1-2\lambda)$. In other words, the perturbative class is associated with the power-law potentials
and a particular exponential potential.

The inflation observables at first order in the Hubble flow parameters correspond to:
\begin{eqnarray}
   r = \frac{16 \lambda}{\Delta N + \lambda} \,, \qquad
   n_s = 1 - \frac{1+2\lambda}{\Delta N + \lambda} \,.
   \label{eq:observables_pert}
\end{eqnarray}
In figure \ref{fig:pert} we can see the restrictions on the parameters $N_*$ and $\lambda$ that
come from the observations of the CMB (the results are summarised in table \ref{tab:parameter_regions}).
From the figure we see that the preferred values for the number of
e-folds are between $23$ and $63$ which includes the standard range between $50$ and $60$.
The values of $\lambda$ accepted at $95\%$ confidence level are smaller than 0.23 which corresponds to
potentials of the form $T^n$ with $n$ smaller than $1.7$, in particular, the potential $T^2$ which
corresponds to $\lambda=0.25$ lies outside the $95\%$ confidence region, as well as the particular case
of the exponential potential, realised for $\lambda = 0.5$.

The field range is given by
\begin{equation} 
  \Delta \ttilde = \frac{3.96 \times 10^3}{\mpl}\sqrt{\frac{\Delta N + \lambda}{\lambda}} \times
  \begin{cases}
     \frac{\lambda}{1-2\lambda}  
     \left[ \sqrt{\frac{\Delta N + \lambda}{\lambda}} - \left( \frac{\Delta N + \lambda}{\lambda} \right)^\lambda
     \right] & 0 < \lambda<\frac{1}{2} 
     \,,\\
     \frac{\sqrt{2\Delta N + 1 }}{4}\log\left( 2\Delta N + 1 \right)
     & \lambda = \frac{1}{2} \,.
  \end{cases}
\end{equation}

\noindent For this class of models the tachyon field can be transformed to a canonical field for the slow-roll period using the transformation (\ref{eq:canonical_transformation}), which gives an excursion of
\begin{equation}
   \Delta \phi = \sqrt{8\lambda} \mpl \left( \sqrt{\Delta N + \lambda} - \sqrt{\lambda} \right) \,.
   \label{eq:canonical_pert}
\end{equation}

In Table \ref{tablota} we compute the field excursion for different models inside the $2\sigma$
region including the best fit values of $\Delta N=37$, $\lambda=0.015$.
One of the main results for this class is that the typical cases
allowed by the observations also present a trans-planckian excursion
in the canonical field.
It is possible however, as seen in Table \ref{tablota}, to obtain sub-planckian excursions but
for very atypical values of the potential, for example writing $n_s=0.97$ and $\Delta \phi=1$
gives us $\lambda=3\times10^{-3}$ which is a very flat potential as it corresponds to
$V\propto T^{0.012}$.

\begin{table}
   \centering
   \begin{tabular}{cccccc}
      \hline
      \hline
      Class & Parameter &\multicolumn{2}{c}{Planck + BAO} & \multicolumn{2}{c}{BK14 + Planck + BAO}
      \\ \hline
       & & Mean & 95\% limits & Mean & 95\% limits
       \\ \cline{3-6}
       \multirow{2}{*}{Perturbative} & $\Delta N$ & 41 & $(25, 64)$ & 41 & $(26, 63)$
       \\ & $\lambda (\times 10^{-2})$ & 11 & $< 35$ & 8 & $<23$

       \\ \hline
       \multirow{2}{*}{Polinomial} & $\Delta N$ & 57 & $> 34$ & 56 & $> 35$
       \\ & $\lambda (\times 10^{-2})$ & 44 & $< 81$ & 39 & $<77$

       \\ \hline
       \multirow{2}{*}{Exponential} & $\Delta N$ & 57 & $> 29$ & 59 & $> 35$  
       \\ & $\lambda (\times 10^{-2})$ & 3.2 & (2.4, 4.0) & 3.1 & $(2.2, 4.0)$

       \\ \hline
       \multirow{2}{*}{Sech} & $\Delta N$ & 69 & $> 61$ & 71 & $> 65$ 
       \\ & $\lambda (\times 10^{-2})$ & 1.0 & $(0.04, 1.6)$ & 1.2 & $(0.5, 1.9)$

   \end{tabular}
   \caption{Marginalized limits of the inflation parameters derived from data of Planck 2015 polarization, BICEP-Keck 2014
   and BAO.}
   \label{tab:parameter_regions}
\end{table}


\subsection{Polynomial class}
\label{Sec:IIIb}
The polynomial class has a {Hubble flow} parameter 
\begin{equation}
\epsilon_1=\frac{\lambda}{N( N^{2\lambda}+1)} \,.
\end{equation}
Therefore, the inflation period for this class ends when this parameter
becomes one, at
\begin{equation}
   \label{Nfinalpol}
N_e(N_e^{2\lambda}+1) = \lambda \,.
\end{equation}
Using the expressions (\ref{classes_basic}), the potential as a function of $T$ reads
\begin{equation}\label{pot_pol}
   \vtilde( \ttilde ) =  
   \begin{cases}
     V_{0}\frac{A\ttilde^{n}}{A\ttilde^{n}+1} & 0<\lambda<\frac{1}{2}
     \,,\\
     \frac{V_0}{2} \left\lbrack 1 +
     \tanh \left(\sqrt{\frac{V_0}{4\mpl^2}} \ttilde \right) \right\rbrack & \lambda=\frac{1}{2}
     \,,\\
     V_{0}\frac{A}{A+\ttilde^{n}} & \lambda>\frac{1}{2},
   \end{cases}
\end{equation}
where $A=\lbrack V_{0}\left(1-2\lambda\right)^{2}/(8\mpl^2\lambda)\rbrack^{2 \lambda/\left(1-2\lambda\right)}$ and 
$n=4 \lambda/|1-2\lambda|$.

The observable parameters of this class are given by:
\begin{eqnarray}
   r = \frac{16 \lambda}{N_*(1+N_*^{2\lambda})} \,, \quad
   n_s =  1 - \frac{1+2\lambda}{N_*} \,,
   \label{eq:pol_observables}
\end{eqnarray}
where $N_*=\Delta N + N_e$ and $N_e$ is the solution to Eq. (\ref{Nfinalpol}).
In figure \ref{fig:pol} we show the restrictions on the parameters $N$ and $\lambda$ as well as its
limits in Table \ref{tab:parameter_regions}. From the figure, we can see that both values are
highly correlated, with the limits only marked by the priors of the model. From 
expressions (\ref{eq:pol_observables}) we note that this class naturally 
produces small tensor perturbations, therefore it is not constrained by
the value of $r$ but only by its prediction on $n_s$.

The field range for this class is given by
\begin{equation}
   \Delta \ttilde = 
   \frac{3.96\times10^3 N_*}{\mpl}
   \begin{cases}
      \frac{1}{1-2\lambda} \left( 1 -
      \left( \frac{N_e}{N_*} \right)^{(1-2\lambda)/2}\right) & \lambda \neq \frac{1}{2} \,,
      \nonumber \\
      \log \left( \frac{2N_*}{\sqrt{3}-1} \right) & \lambda = \frac{1}{2} \,.
   \end{cases}
   \label{}
\end{equation}

\noindent Equivalently, the canonical field associated with this tachyon field is given by the integral
\begin{equation}
\Delta \phi = \mpl \sqrt{2\lambda} \int_{N_e}^{N_*} N^{-1}(N^{2\lambda} + 1)^{-1}dN \,.
\end{equation}

We find that the excursion of the associated canonical scalar field is trans-planckian, except for the smallest values of $\lambda$ and $\Delta N$ (within the observational bounds). This means that the model may prove incomplete since string theory corrections
are required in order to have a dynamically stable model. The values of $(\lambda,\Delta N)$ with an associated small $\Delta \phi$
result on atypical potentials as summarised in Table \ref{tablota}.

\subsection{Exponential class}
\label{Sec:IIIc}
The exponential class corresponds to one in which the Hubble flow parameter is given by
\begin{equation}\label{expfirstepsilon}
   \epsilon_{1} =\frac{\lambda}{2\left( e^{\lambda N}+1\right)} \,,
\end{equation}

\noindent for $\lambda$ positive.

We note that for this class the maximum value of the Hubble flow parameter corresponds
to $\lambda/4$ and the most preferred values of $\lambda$
given the observations (Table \ref{tab:parameter_regions}) correspond to
$\lambda<0.04$. This means that $\epsilon_1$
never reaches a value close to unity, therefore inflation cannot reach an end. One possibility
for this model is to have an auxiliary field
that couples to the inflaton in order to end inflation as in the Hybrid
Inflation scenario \cite{Linde:1993cn}. However we shall not analyze here the particular
mechanism that ends inflation and we simply assume that inflation ends at $N_e=0$ \footnote{
A non zero value of $N_e$ may change the area occupied in the ($n_s,r$) space.}.

The potential for this class as a function of the tachyon field reads
\begin{equation}
   \label{pot_expfirst}
   \vtilde = \frac{V_0}{1 + A \ttilde ^2} \,,
\end{equation}
where $A = \lambda V_0/4\mpl^2$. The observable parameters are in this case:
\begin{equation}
   n_s = 1-\lambda\,, \qquad r = \frac{8 \lambda}{1+e^{\lambda \Delta N}} \,,
\end{equation}

\noindent and we can immediately deduce that $\lambda$ will be restricted to small values due to the almost scale-invariant spectrum observed by Planck.

The field excursion obtained for this class is determined by the following expression
\begin{equation}
   \Delta \ttilde = \frac{3.96\times10^3}{\mpl \lambda}
   (e^{\lambda \Delta N / 2} - 1) \,,
\end{equation}
while we can use the equation (\ref{eq:canonical_transformation})
in order to obtain an expression for the canonical field excursion
associated with this model as
\begin{equation}
   \Delta \phi = \frac{2\mpl}{\sqrt{\lambda}}
   \log\left[ \left( \sqrt{1+e^{-\lambda \Delta N}} - e^{-\lambda \Delta N / 2} \right)
   \left( \sqrt{2} + 1 \right) \right] \,.
   \label{expdeltaphi}
\end{equation}
In Figure \ref{fig:exp} and Table \ref{tab:parameter_regions}
we read the allowed values of $\lambda$ 
between $0.022$ and $0.04$. It can be checked easily from equation
(\ref{expdeltaphi}) that for those values of $\lambda$ a sub-planckian
excursion in the canonical field is only possible if $\Delta N<9.8$ (or
worse, if $\lambda=0.04$, $\Delta N<7.3$ is required). This means that
the exponential class of models cannot solve the problems of the classical
cosmology and simultaneously have a sub-planckian field excursion.

If we ignore the trans-planckian excursion then we have $\Delta N$ valid
for a fairly wide range of possible values, between 53 and 75, which
contain the usual range between 50 and 60, and which makes the model compatible with an
energy scale between the GUT scale and the Planck scale. We can also see from figure \ref{fig:exp} that in
this case the values of $\lambda$ and $\Delta N$ are uncorrelated and can
be determined independently.

\subsection{$V \propto \sech(T)$ class}

In this section we will study a class of potentials inspired in string theory realizations \cite{Daniel,sechinflation} which
are potentials proportional to the hyperbolic secant function.
This class of potentials come from a Hubble flow parameter of the form \footnote{in our previous work this class of models were labelled as the second exponential class}
\begin{equation}
   \epsilon_1 = \lambda \csch (2\lambda N) \,,
\end{equation}

\noindent Its associated potential is given by
\begin{equation}\label{pot_sech}
   \vtilde (\ttilde) = V_{0} \sech \left( \frac{\sqrt{V_0 \lambda}}{\mpl} \ttilde \right) \,.
\end{equation}

\noindent The inflationary period for this model ends at $N_e=\sinh^{-1}(\lambda)/2\lambda$
which for the values of $\lambda$ favored by the observations is approximately
equal to $1/2$ (see Table \ref{tab:parameter_regions}). Also, the observables related to this class are given by
\begin{eqnarray}
   r = 16\lambda \csch(\lambda (2\Delta N +1 )\,, \qquad
   n_s = 1- 2 \lambda \coth(\lambda (\Delta N + 1/2)) \,.
   \label{sechobs}
\end{eqnarray}

 The field excursion for this class is determined by the following expression
\begin{equation}
   \Delta \ttilde = \frac{1.98\times 10^3}{\mpl}
   \frac{\sinh (\lambda (\Delta N + 1/2))}{\lambda}
   \log\left[ \frac{\tanh(\lambda(2\Delta N +1)/4)}{\tanh(\lambda/4)} \right] \,,
\end{equation}
for the tachyon field while the associated canonical field from equation
(\ref{eq:canonical_transformation}) has an excursion given by
\begin{equation}\label{sech_deltaphi}
   \Delta \phi = \sqrt{2\lambda\mpl^2}
   \int_{N_e}^{N_*} \sqrt{\csch(2\lambda N)} dN \,.
\end{equation}

For this class we see in Figure \ref{fig:sech} and in 
Table \ref{tab:parameter_regions} that the observations prefer the largest possible
values of $\Delta N$ allowed by the priors. For example, if we evaluate the observables at the mean value of the r-distribution $(n_s = 0.965, r = 0.048)$ \cite{Array:2015xqh}, Eq.~\eqref{sechobs} yields $(\lambda = 0.014,\Delta N = 82)$. The large value of $\Delta N$ implies the need of a period of
stiff fluid domination (sometimes referred as kination) between the end of inflation and nucleosynthesis,
as noted in Ref. \cite{Liddle:2003as}. This fact introduces a new level of complexity to
this popular model. We avoid these large values in our analysis of data. 
In Table~\ref{tablota} we obtained the results for the mean values of the parameters according
to the BK14 observations. The amplitude of tensor perturbations barely enter into the valid range allowed
by the observations and that corresponds to high values of $\Delta N$ as the
expression (\ref{sechobs}) can be approximated for small $\lambda$ to $r\approx 8 /\Delta N$.
If we compute the canonical field excursion \ref{sech_deltaphi} for the allowed
values of Table \ref{tab:parameter_regions} we can see that $\Delta \phi>13$
meaning that the excursion is trans-planckian, which is a theoretical limitation for this class.

\begin{table}[bbp]
  \centering
\begin{tabular}{cccccccc}
  \hline
  \hline
  Class & $V(T)\propto$ & $\lambda$ & $\Delta N$ & $\Delta \ttilde \mpl$ & $\Delta \phi / \mpl$ & $r$ & $n_s$\\
  \hline
  \multirow{4}{*}{Perturbative}&
  $T^{0.06}$& $0.015$ & 37 & $1\times10^5$ & 2 & 0.006 & 0.972 \\
  &$\sqrt{T}$& $0.1$ & 46 & $2\times10^5$ & 5.7 & 0.03 & 0.974 \\
  & $T$& $1/6$ & 50 & $2\times10^5$ & 7.7 & 0.05 & 0.973 \\
  & $T^{0.004}$& $0.001$ & 50 & $2\times10^5$ & 0.6 & $3\times10^{-4}$ & 0.979 \\
  \hline
  \multirow{4}{*}{Polynomial}&
 $T^{2}/(A T^{2}+1)$ & $1/4$ & 45  & $2\times10^5$ & $4.1$ & $0.01$ & $0.967$ \\
  &$ 1 + \tanh (AT)$ & $1/2$ & $60$ & $1\times 10^{6}$ & $4.0$&$2\times 10^{-3}$&$0.967$ \\
  &$1/(A+T^{6})$ & $3/4$ & 75 & $1 \times 10^{6}$ & $3.3$ & $2\times 10^{-4}$ &$0.967$\\
  &$T^{0.04}/(A T^{0.04}+1)$ & $10^{-2}$ & 30  & $10^5$ & $0.94$ & $2\times10^{-3}$ & $0.966$ \\
  \hline
  \multirow{2}{*}{Exponential}&
  $(1 + AT ^2)^{-1}$ & $0.03$ & 59 & $2 \times 10^{5}$ & $5.5$  & $0.03$&$0.970$\\
  &$(1 + AT^2)^{-1}$ & $0.03$ & 35 & $9 \times 10^{4}$ & $3.7$  & $0.06$&$0.970$\\
  \hline
  $V\propto \sech(T)$&
  $\sech (A T)$ & $0.012$ & 71 & $8 \times 10^{5}$ & $14.7$  & $0.07$&$0.965$\\
\end{tabular}
\bigskip
\caption{Field range and observational predictions for selected potentials within each of the studied universality classes.
All potentials included are derived from values within the 95\% confidence level of the Bayesian parameter estimation analysis (see Table~\ref{tab:parameter_regions}).
For the perturbative class the best fit and two simple potentials are included which have
a trans-planckian excursion. The last potential of this class has a sub-planckian
excursion.
For the polynomial class a range of potentials fit
well within the observational bounds, but the canonical
field excursion associated with the models is trans-planckian except for those models with
the smallest $\lambda$ and $\Delta N$ within the bounds. For the exponential potential we have
two cases with different $\Delta N$ but the same $\lambda$ as the latter is very well constrained
by the observations. All the models within this class that satisfy the observational bounds are trans-planckian.
For the $V \propto \sech(T)$ class, only the mean value is shown, which corresponds to a trans-planckian excursion with
a tensor-to-scalar ratio at the maximum value allowed by the observations.}
\label{tablota}
\end{table}

\section{Bayesian model selection}\label{Sec:modelselection}

From the plots in Figure \ref{fig:obs_nlam} and Table \ref{tab:parameter_regions} we see that 
the expected values for $\lambda$ and $N^{*}$ are highly dependent on the particular universality class.
This expected since the functional dependence on the parameters of the primordial power spectrum
($n_s$, $r$, $n_{sk}$, $n_t$) is different for each class. The results are obtained assuming in each
case that the correct model of inflation is that of the particular class, therefore they only tell
us about the preferred values for the parameters within
each model but not whether the model is compatible with the
observations or not.

\begin{figure}[h]
   \centering
   \psfig{figure=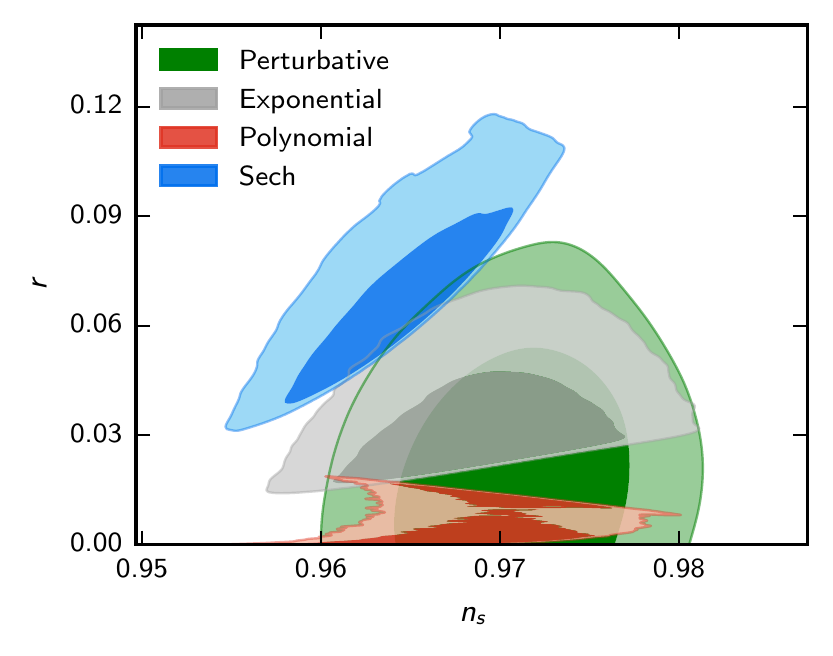}
   \caption{Marginalized confidence regions for each of the universality classes in the $r$ vs
   $n_s$ plane, from BK14, Planck and BAO data, as described in the main text. Each class produces different values of $r$ and $n_s$ as functions
of $\lambda$ and $N_{*}$, which are given flat priors. This produces different confidence
regions given the data. Most of the surface sampled by the $V \propto \sech(T)$ model 
lies outside the area preferred by observations, while the exponential model is well inside it. The Bayesian model comparison indicates which of these tachyon field classes is preferred by observations.}
   \label{fig:all_models}
\end{figure}

Despite each of the models having a region of maximum probability
it might not correspond to a model compatible enough with the data, as
can be seen from the figure \ref{fig:all_models}. In particular, we can observe that the 
prediction for the hyperbolic secant model corresponds to a higher value of the tensor perturbations
than that of the other classes.
Equation  (\ref{sechobs}) indicates that small values of $r$ require large values of $\Delta N$ outside the assumed range.
This means that the most likely values of $\Delta N$ given the data fall out of range in this particular class of the Tachyon model. The parameter estimation within the imposed range is blind to this fact because the probabilities are not normalised, but the amplitude
of the probability (relative to that of other classes) is calculated in the Bayesian Model Selection analysis  as shown in this section. 

The parameter estimation analysis provides a measurement of
the posterior probability given by the Bayes theorem as
\begin{equation}
   \mathcal{P}_M (\theta) = \frac{\mathcal{L}_M (\theta)\pi_M(\theta)}
   {\mathcal{Z}_M} \,,
   \label{}
\end{equation}
where $\theta$ is the vector of the parameters of the model $M$, $\mathcal{L}_M$
is the likelihood which is a function of the data, $\pi_M$ is the prior and
$\mathcal{Z}$ is the evidence which functions as a normalization factor. In order
to do the parameter estimation within a particular model, as it is done in
the figures \ref{fig:obs_nlam} and \ref{fig:all_models}, the evidence is not
considered and we plot an unnormalized posterior function. However, a Bayesian model comparison demands an estimation of the evidence in order to obtain the probability of each class of models, and this is given by
\begin{equation}
   \mathcal{Z}_M =\int_\Omega \mathcal{L}(\theta) \pi(\theta) d\theta \,,
   \label{zeta-M}
\end{equation}

\noindent where $\Omega$ is the parameter-space of $\theta$.
This integral is modulated by the priors which pick only the region allowed by the analysis in Sec.~\ref{Sec:III}. As mentioned earlier, this may exclude the maximum likelihood region for a specific class and provide a smaller evidence. 
To explicitly compute this integral, we use the code POLYCHORD \cite{polychord}
which implements a nested sampling algorithm within the COSMOMC framework. We compared
the evidence of each class with the one obtained for 
a model where the scalar spectral index $n_s$ and the tensor to scalar ratio $r$
were considered as independent variables with flat priors; this is our reference model for which no specific inflationary model is imposed.
The set of data used corresponds to BK14 + Planck + BAO as in the parameter estimation analysis. The results are summarized in Table~\ref{tab:evid}, where we use the Bayes factor $B_i=Z_i/Z_{\rm ref}$, with $Z_{\rm ref}$ the evidence of 
the reference model. 

Using the scale suggested by Kass and Raftery in \cite{kassraft},  a \textit{substantial} evidence is obtained against the $V\propto \sech(T)$  class. Meanwhile, the evidence of the exponential class is around 1 (similar to that of the reference model). Finally, both the polynomial and perturbative classes show \textit{substantial} evidence favouring both in comparison with the reference model.

\begin{table}
   \centering
   \begin{tabular}{ccc}
      \hline
      \hline
      Class & $\log(B)$ & B
      \\ \hline
      Polynomial & 2.4 & 11 \\
      Perturbative & 1.4 & 4.1 \\
      Exponential & -0.06 & 0.94 \\
      Sech & -1.4 & 0.25 
   \end{tabular}
   \caption{Bayes factor between the different classes as compared to a reference model with $n_s$ and $r$
   as free parameters with flat priors. The evidence was computed using the POLYCHORD code on the BK14+PLANCK+BAO joint dataset.}
   \label{tab:evid}
\end{table}

\section{Discussion and concluding remarks}\label{Sec:V}

In this paper we have analysed an inflation scenario driven by a
tachyon condensate with the goal of contrasting the several universality classes that arise from the large-$N$ formalism \cite{Nan} against observational data and theoretical considerations.

A first approximation to quantify the excursion of the tachyonic
field is to follow a similar path to that taken by Lyth in Ref. \cite{Lyth:1996im}
using the COBE normalization to obtain a bound on the canonical
field excursion. The tachyon field follows a modified excursion with respect to the canonical inflaton.
We have shown that the tachyon excursion follows the expression (\ref{firstbound}) that is only a function of the number of e-folds and
not of the amplitude of the tensor perturbations, this expression was also obtained in Ref. \cite{Fei:2017fub}
and implies that for
$N \sim 10^2$ the excursion for the tachyon field is of order $10^5 \mpl^{-1}$.
Since the units of the tachyon field are $\mpl^{-1}$, we cannot assess directly a sub-planckian or trans-planckian excursion. 
In order to evaluate the relevance of quantum corrections to the classical field analysis, it is customary to make a transformation
to a canonical field which then can be quantized. We have therefore transformed the excursion to the canonical field following Eq.~(\ref{eq:canonical_transformation}) for each of the universality classes. The result is given analytically for the perturbative
and the exponential classes (Eqs.~\eqref{eq:canonical_pert} and \eqref{expdeltaphi} respectively) and computed numerically for the polynomial and $V \propto \sech(T)$ classes.

Regarding cosmological observations, we constrain each of the universality classes performing a parameter estimation analysis on each of them. 
We employ the CMB polarization data from Planck and BK14, both of 
which provide an upper bound to the tensor-to-scalar ratio $r$, together with 
the CMB temperature data from Planck that gives a very precise value of
the spectral index of the scalar perturbations $n_s$. In figure \ref{fig:obs_nlam} we present the posterior distribution for the pair of parameters
($\lambda$, $\Delta N$) for each of the considered universality classes. 
For the perturbative class we see that the simple potentials $V\propto T^n$
with $n\ge 2$ lie outside the 95\% confidence region. Potentials proportional to $T$ and
$\sqrt{T}$ lie inside the region favoured by the observations, but
with a trans-planckian canonical field excursion (see Tabla~\ref{tablota}). It is possible
to obtain models allowed by the observations and with sub-planckian excursions, but with
smaller values on the exponent of the potential, as shown in Table~\ref{tablota} for $V\propto T^{0.004}$.

For the $V \propto \sech(T)$ model and the exponential model, the production of gravitational
waves has a minimum which may be excluded by future probes of the CMB polarization if no gravitational
wave signal is detected.

On the other hand, the polynomial class has the generic property of producing
a small value of $r$ (see Figure~\ref{fig:all_models}). For this reason, the only effective constraint on the model parameters 
is the spectral index which, together with the priors, yields the posterior distribution
shown in Figure~\ref{fig:pol}. In this sense, our results are consistent with similar parametrisations of the non-canonical 
model~\cite{Binetruy:2016hna}. However, as in the case of the perturbative class, the
simplest potentials that lie within the observational bounds (for example, $V\propto T^2/(1+AT^2)$)
show the problem of a trans-planckian excursion in the associated 
canonical potential. There is, however, a region of the parameter space with
sub-planckian excursions that is also favoured by the observations, with the compromise of small exponents, e.g. $V\propto T^{0.04}/(1+AT^{0.04})$ (see Table~\ref{tablota}). 

For the exponential class a significant problem that arises is the lack of a natural
exit from inflation since $\epsilon_1$ never reaches values of order one. Even if 
one finds an alternative mechanism to end inflation, the parameter space
favoured by the observations has $\Delta \phi>\mpl$ which means that the classical treatment
is incomplete until the string theory corrections are taken into account to ensure they are
dynamically controlled.

As shown in Sec.~\ref{Sec:V}, for the $V \propto \sech(T)$ class, the excessive production of tensor perturbations reduces the relative evidence of the model (see Table~\ref{tab:evid}). In the parameter estimation analysis this is not a problem as the model is assumed valid in order to compute the posterior. As happens with the exponential class, all of the sampled region in parameter-space corresponds to models with a trans-planckian excursion in the associated canonical field. An additional argument against this class of potentials is that the sector compatible with the observational bounds on $r$ requires a stiff fluid phase of evolution after inflation. This is because this class yields a large value of $\Delta N$. 
 These arguments deem the $V \propto \sech(T)$ class unlikely.

In conclusion, our results show that in the context of tachyon Inflation, the original $V \propto \sech(T)$ potential is disfavoured both by observations and the theoretical preference of sub-panckian excursions. The same goes for the exponential class of models. On the contrary, we argue that the polynomial and perturbative classes of models are statistically and theoretically favoured if the potentials are sufficiently flat. This is summarised in Tables~\ref{tablota} and~\ref{tab:evid} and shows consistency with studies of several realisations of the canonical field potential \cite{vennin:papers}.

\acknowledgments

We gratefully acknowledge support from \textit{Programa de Apoyo a Proyectos de Investigaci\'on e Innovaci\'on
Tecnol\'ogica} (PAPIIT) UNAM, IA-103616, \textit{Observables en Cosmolog\'{\i}a Relativista} and SEP-CONACYT grant No. 239639. 
All authors thank SNI-CONACYT for partial support.



\begin{thebibliography}{99}

\bibitem{inflation}
A.H.~Guth, ``The Inflationary Universe: A possible solution to the horizon and flatness problems,''
Phys.\ Rev.\ \textbf{D23} (1981) 347. 
A.D~Linde,
 ``A New Inflationary Universe Scenario: A Possible Solution of the Horizon, Flatness, Homogeneity, Isotropy and Primordial Monopole Problems,''
Phys.\ Lett.\ \textbf{B108} (1982) 389. 
A.~Albrecht and P.J.~Steinhardt,
`` Cosmology for Grand Unified Theories with Radiatively Induced Symmetry Breaking,''
Phys.\ Rev.\ Lett.\ \textbf{48} (1982) 1220. 
D.~H.~Lyth and A.~Riotto,
 ``Particle physics models of inflation and the cosmological density perturbation,''
 Phys.\ Rept.\  {\bf 314} (1999) 1 [hep-ph/9807278].
  A.~A.~Starobinsky,
  ``A New Type of Isotropic Cosmological Models Without Singularity,''
  Phys.\ Lett.\  {\bf 91B} (1980) 99.
  doi:10.1016/0370-2693(80)90670-X.

\bibitem{non-canonical}
  C.~Armendariz-Picon, T.~Damour and V.~F.~Mukhanov,
  ``k - inflation,''
  Phys.\ Lett.\ B {\bf 458} (1999) 209
  doi:10.1016/S0370-2693(99)00603-6
  [hep-th/9904075].
  J.~Garriga and V.~F.~Mukhanov,
  ``Perturbations in k-inflation,''
  Phys.\ Lett.\ B {\bf 458} (1999) 219
  doi:10.1016/S0370-2693(99)00602-4
  [hep-th/9904176].
 \bibitem{sen:todas}
 A.~Sen,
 ``BPS D-branes on nonsupersymmetric cycles,''
 JHEP {\bf 9812} (1998) 021
 [hep-th/9812031].
A.~Sen,
``Supersymmetric world volume action for nonBPS D-branes,''
JHEP {\bf 9910} (1999) 008
[hep-th/9909062].
 A.~Sen,
``Tachyon condensation on the brane anti-brane system,''
 JHEP {\bf 9808} (1998) 012
[hep-th/9805170].
A.~Sen, ``Tachyon matter,''
JHEP {\bf 0207} (2002) 065
[hep-th/0203265].
 A.~Sen,
 ``Rolling tachyon,''
JHEP {\bf 0204} (2002) 048
[hep-th/0203211].
 A.~Sen,
``Field theory of tachyon matter,''
 Mod.\ Phys.\ Lett.\ A {\bf 17} (2002) 1797
[hep-th/0204143].
J.~De-Santiago and J.~L.~Cervantes-Cota,
  ``Generalizing a Unified Model of Dark Matter, Dark Energy, and Inflation with Non Canonical Kinetic Term,''
  Phys.\ Rev.\ D {\bf 83} (2011) 063502
  doi:10.1103/PhysRevD.83.063502
  [arXiv:1102.1777 [astro-ph.CO]].
\bibitem{Daniel}
 D.~Cremades, F.~Quevedo and A.~Sinha, ``Warped tachyonic inflation in type IIB flux compactifications and the open-string completeness conjecture,'' JHEP {\bf 0510} (2005) 106 [hep-th/0505252].

\bibitem{tachyon1}
 D.~A.~Steer and F.~Vernizzi,
``Tachyon inflation: Tests and comparison with single scalar field inflation,''
Phys.\ Rev.\ D {\bf 70}, 043527 (2004)
[hep-th/0310139].

\bibitem{Li:2013cem}
S.~Li and A.~R.~Liddle,
``Observational constraints on tachyon and DBI inflation,''
JCAP {\bf 1403}, 044 (2014)
[arXiv:1311.4664 [astro-ph.CO]].

\bibitem{otrostaquiones}
  M.~R.~Garousi,
  ``On shell S matrix and tachyonic effective actions,''
  Nucl.\ Phys.\ B {\bf 647} (2002) 117
  doi:10.1016/S0550-3213(02)00903-3
  [hep-th/0209068].
  M.~R.~Garousi,
  ``Tachyon couplings on nonBPS D-branes and Dirac-Born-Infeld action,''
  Nucl.\ Phys.\ B {\bf 584} (2000) 284
  doi:10.1016/S0550-3213(00)00361-8
  [hep-th/0003122].
 E.~A.~Bergshoeff, M.~de Roo, T.~C.~de Wit, E.~Eyras and S.~Panda,
  ``T duality and actions for nonBPS D-branes,''
  JHEP {\bf 0005} (2000) 009
  doi:10.1088/1126-6708/2000/05/009
  [hep-th/0003221].
  J.~Kluson,
  ``Proposal for nonBPS D-brane action,''
  Phys.\ Rev.\ D {\bf 62} (2000) 126003
  doi:10.1103/PhysRevD.62.126003
  [hep-th/0004106].
  M.~Sami, P.~Chingangbam and T.~Qureshi,
  ``Aspects of tachyonic inflation with exponential potential,''
  Phys.\ Rev.\ D {\bf 66} (2002) 043530
  [hep-th/0205179].
  M.~R.~Garousi,
  ``Slowly varying tachyon and tachyon potential,''
  JHEP {\bf 0305} (2003) 058
  doi:10.1088/1126-6708/2003/05/058
  [hep-th/0304145].
  K.~Nozari and N.~Rashidi,
  ``Some Aspects of Tachyon Field Cosmology,''
  Phys.\ Rev.\ D {\bf 88} (2013) 2,  023519
  [arXiv:1306.5853 [gr-qc]].

\bibitem{Nan}
N.~Barbosa-Cendejas, J.~De-Santiago, G.~German, J.~C.~Hidalgo and R.~R.~Mora-Luna,
  ``Tachyon inflation in the Large-$N$ formalism,''
  JCAP {\bf 1511} (2015) 020
  doi:10.1088/1475-7516/2015/11/020
  [arXiv:1506.09172 [astro-ph.CO]].

\bibitem{nformalism}
D.~Boyanovsky, H.~J.~de Vega and N.~G.~Sanchez,
``Clarifying Inflation Models: Slow-roll as an expansion in $1/N_{e-folds}$''
Phys.\ Rev.\ D {\bf 73} (2006) 023008
[astro-ph/0507595].
  V.~Mukhanov,
 ``Quantum Cosmological Perturbations: Predictions and Observations,''
  Eur.\ Phys.\ J.\ C {\bf 73} (2013) 2486
   [arXiv:1303.3925 [astro-ph.CO]].
D.~Roest,
``Universality classes of inflation,''
 JCAP {\bf 1401} (2014) 01,  007
 [arXiv:1309.1285 [hep-th]].
  J.~Garcia-Bellido, D.~Roest, M.~Scalisi and I.~Zavala,
 ``Can CMB data constrain the inflationary field range?,''
  JCAP {\bf 1409} (2014) 006
 [arXiv:1405.7399 [hep-th]].
  J.~Garcia-Bellido and D.~Roest,
 ``Large-$N$ running of the spectral index of inflation,''
  Phys.\ Rev.\ D {\bf 89} (2014) 10,  103527
 [arXiv:1402.2059 [astro-ph.CO]].
  Q.~Gao, Y.~Gong and Q.~Fei,
  ``Constant-roll tachyon inflation and observational constraints,''
  arXiv:1801.09208 [gr-qc].


\bibitem{Fei:2017fub} 
  Q.~Fei, Y.~Gong, J.~Lin and Z.~Yi,
  JCAP {\bf 1708}, no. 08, 018 (2017)
  doi:10.1088/1475-7516/2017/08/018
  [arXiv:1705.02545 [gr-qc]].
\bibitem{Garcia-Bellido:2014wfa}
  J.~Garcia-Bellido, D.~Roest, M.~Scalisi and I.~Zavala,
  ``Lyth bound of inflation with a tilt,''
  Phys.\ Rev.\ D {\bf 90} (2014) 12,  123539
  [arXiv:1408.6839 [hep-th]].

\bibitem{Baumann:2009ds}
D.~Baumann,
 ``Inflation,'
doi:10.1142/97898143271830010
 arXiv:0907.5424 [hep-th].

\bibitem{Adam:2015rua}
  R.~Adam {\it et al.}  [Planck Collaboration],
  ``Planck 2015 results. I. Overview of products and scientific results,''
  [arXiv:1502.01582 [astro-ph.CO]].

\bibitem{Array:2015xqh}
  P.~A.~R.~Ade {\it et al.} [BICEP2 and Keck Array Collaborations],
  ``Improved Constraints on Cosmology and Foregrounds from BICEP2 and Keck Array Cosmic Microwave Background Data with Inclusion of 95 GHz Band,''
  Phys.\ Rev.\ Lett.\  {\bf 116} (2016) 031302
  doi:10.1103/PhysRevLett.116.031302
  [arXiv:1510.09217 [astro-ph.CO]].

\bibitem{baotodos}
  F.~Beutler {\it et al.},
  ``The 6dF Galaxy Survey: Baryon Acoustic Oscillations and the Local Hubble Constant,''
  Mon.\ Not.\ Roy.\ Astron.\ Soc.\  {\bf 416}, 3017 (2011)
  doi:10.1111/j.1365-2966.2011.19250.x
  [arXiv:1106.3366 [astro-ph.CO]].
  A.~J.~Ross, L.~Samushia, C.~Howlett, W.~J.~Percival, A.~Burden and M.~Manera,
  ``The clustering of the SDSS DR7 main Galaxy sample – I. A 4 per cent distance measure at $z = 0.15$,''
  Mon.\ Not.\ Roy.\ Astron.\ Soc.\  {\bf 449} (2015) no.1,  835
  doi:10.1093/mnras/stv154
  [arXiv:1409.3242 [astro-ph.CO]].
  L.~Anderson {\it et al.} [BOSS Collaboration],
  ``The clustering of galaxies in the SDSS-III Baryon Oscillation Spectroscopic Survey: baryon acoustic oscillations in the Data Releases 10 and 11 Galaxy samples,''
  Mon.\ Not.\ Roy.\ Astron.\ Soc.\  {\bf 441} (2014) no.1,  24
  doi:10.1093/mnras/stu523
  [arXiv:1312.4877 [astro-ph.CO]].

\bibitem{alfredo}
  N.~Barbosa-Cendejas, R.~Cartas-Fuentevilla, A.~Herrera-Aguilar, R.~R.~Mora-Luna and R.~da Rocha,
  JCAP {\bf 1801} (2018) no.01,  005
  doi:10.1088/1475-7516/2018/01/005
  [arXiv:1709.09016 [hep-th]].

  G.~Germán, A.~Herrera-Aguilar, A.~M.~Kuerten, D.~Malagon-Morejon and R.~da Rocha,
  JCAP {\bf 1601} (2016) no.01,  047
  doi:10.1088/1475-7516/2016/01/047
  [arXiv:1508.03867 [hep-th]].

  G.~German, A.~Herrera-Aguilar, D.~Malagon-Morejon, R.~R.~Mora-Luna and R.~da Rocha,
  JCAP {\bf 1302} (2013) 035
  doi:10.1088/1475-7516/2013/02/035
  [arXiv:1210.0721 [hep-th]].


\bibitem{Schwarz:2001vv}
  D.~J.~Schwarz, C.~A.~Terrero-Escalante and A.~A.~Garcia,
  ``Higher order corrections to primordial spectra from cosmological inflation,''
  Phys.\ Lett.\ B {\bf 517} (2001) 243;
 [astro-ph/0106020].
 
 \bibitem{Binetruy:2014zya}
  P.~Binetruy, E.~Kiritsis, J.~Mabillard, M.~Pieroni and C.~Rosset,
  ``Universality classes for models of inflation,''
  JCAP {\bf 1504} (2015) no.04,  033
  doi:10.1088/1475-7516/2015/04/033
  [arXiv:1407.0820 [astro-ph.CO]].

\bibitem{Lyth:1996im} 
  D.~H.~Lyth,
  ``What would we learn by detecting a gravitational wave signal in the cosmic microwave background anisotropy?,''
  Phys.\ Rev.\ Lett.\  {\bf 78}, 1861 (1997)
  doi:10.1103/PhysRevLett.78.1861
  [hep-ph/9606387].

\bibitem{Lidsey:1995np}
  J.~E.~Lidsey, A.~R.~Liddle, E.~W.~Kolb, E.~J.~Copeland, T.~Barreiro and M.~Abney,
 ``Reconstructing the inflation potential : An overview,''
 Rev.\ Mod.\ Phys.\  {\bf 69} (1997) 373
   [astro-ph/9508078].

\bibitem{macorra}
A. de la Macorra, U. Filobello and G. German
``The Mass, Normalization and Late Time behavior of the Tachyon Field,''
Phys.\ Lett.\ B {\bf 635} 355 - 363 (2006)

\bibitem{Lyth:2009zz}
  D.~H.~Lyth and A.~R.~Liddle,
  ``The primordial density perturbation: Cosmology, inflation and the origin of structure,''
  Cambridge, UK: Cambridge Univ. Pr. (2009) 497 p

\bibitem{Liddle:2003as}
  A.~R.~Liddle and S.~M.~Leach,
  ``How long before the end of inflation were observable perturbations produced?,''
  Phys.\ Rev.\ D {\bf 68} (2003) 103503
  doi:10.1103/PhysRevD.68.103503
  [astro-ph/0305263].

\bibitem{Lewis:2002ah}
  A.~Lewis and S.~Bridle,
  ``Cosmological parameters from CMB and other data: A Monte Carlo approach,''
  Phys.\ Rev.\ D {\bf 66}, 103511 (2002)
  [astro-ph/0205436].
   
\bibitem{Lewis:1999bs}
  A.~Lewis, A.~Challinor and A.~Lasenby,
  ``Efficient computation of CMB anisotropies in closed FRW models,''
  Astrophys.\ J.\  {\bf 538} (2000) 473
  doi:10.1086/309179
  [astro-ph/9911177].

\bibitem{Linde:1993cn} 
  A.~D.~Linde,
  ``Hybrid inflation,''
  Phys.\ Rev.\ D {\bf 49}, 748 (1994)
  doi:10.1103/PhysRevD.49.748
  [astro-ph/9307002].

\bibitem{sechinflation}
  N.~D.~Lambert, H.~Liu and J.~M.~Maldacena,
  ``Closed strings from decaying D-branes,''
  JHEP {\bf 0703}, 014 (2007)
  doi:10.1088/1126-6708/2007/03/014
  [hep-th/0303139].
  D.~Kutasov and V.~Niarchos,
  ``Tachyon effective actions in open string theory,''
  Nucl.\ Phys.\ B {\bf 666} (2003) 56
  doi:10.1016/S0550-3213(03)00498-X
  [hep-th/0304045].
  K.~Okuyama,
  ``Wess-Zumino term in tachyon effective action,''
  JHEP {\bf 0305} (2003) 005
  doi:10.1088/1126-6708/2003/05/005
  [hep-th/0304108].
G.~W.~Gibbons,
``Cosmological evolution of the rolling tachyon,''
Phys.\ Lett.\ B {\bf 537} (2002) 1
 doi:10.1016/S0370-2693(02)01881-6
[hep-th/0204008].
J.  Raeymaekers,  ``Tachyonic inflation in a warped string background,''
  JHEP {\bf 0410} (2004) 057
  doi:10.1088/1126-6708/2004/10/057
  [hep-th/0406195].

\bibitem{polychord}
  W.~J.~Handley, M.~P.~Hobson and A.~N.~Lasenby,
  ``PolyChord: nested sampling for cosmology,''
  Mon.\ Not.\ Roy.\ Astron.\ Soc.\  {\bf 450} (2015) no.1,  L61
  doi:10.1093/mnrasl/slv047
  [arXiv:1502.01856 [astro-ph.CO]].
  W.~J.~Handley, M.~P.~Hobson and A.~N.~Lasenby,
  ``PolyChord: nested sampling for cosmology,''
  Mon.\ Not.\ Roy.\ Astron.\ Soc.\  {\bf 453} (2015) no.4, 4384
  doi:10.1093/mnras/stv1911
  [arXiv:1502.01856 [astro-ph.CO]].

\bibitem{kassraft}
   Robert E. Kass and Adrian E. Raftery,
   Journal of the American Statistical Association {\bf 90} (1995), 430
   doi:10.1080/01621459.1995.10476572

\bibitem{Binetruy:2016hna}
  P.~Bin\'etruy, J.~Mabillard and M.~Pieroni,
  ``Universality in generalized models of inflation,''
  JCAP {\bf 1703} (2017) no.03,  060
  doi:10.1088/1475-7516/2017/03/060
  [arXiv:1611.07019 [gr-qc]].

\bibitem{vennin:papers}
 J.~Martin, C.~Ringeval and V.~Vennin,
  ``Encyclop{\ae}dia Inflationaris,''
  Phys.\ Dark Univ.\  {\bf 5-6} (2014) 75
  doi:10.1016/j.dark.2014.01.003
  [arXiv:1303.3787 [astro-ph.CO]].
 J.~Martin, C.~Ringeval, R.~Trotta and V.~Vennin,
  ``The Best Inflationary Models After Planck,''
  JCAP {\bf 1403} (2014) 039
  doi:10.1088/1475-7516/2014/03/039
  [arXiv:1312.3529 [astro-ph.CO]].

\end{thebibliography}
\end{document}